\documentclass[doublecol]{epl2}
\usepackage{graphicx,psfrag,color,hyperref}
\usepackage{amssymb,latexsym,mathrsfs,amsmath,mathrsfs}
\def\be{\begin{equation}}
\def\ee{\end{equation}}
\def\bea{\begin{eqnarray}}
\def\eea{\end{eqnarray}}
\def\th{\theta}
\newcommand{\dd}{\mbox{d}}

\usepackage{ulem}
\begin{document}

\title{Dynamics of coupled oscillator systems in presence
of a local potential}
\author{Alessandro Campa\inst{1} and Shamik Gupta\inst{2}}
\shortauthor{A. Campa and S. Gupta} 
\institute{\inst{1} Complex Systems and Theoretical Physics Unit, Health and Technology Department,
Istituto Superiore di Sanit\`{a}, and INFN Roma1, Viale Regina Elena 299, 00161 Roma, Italy \\\inst{2} Max Planck Institute for the Physics of Complex
Systems, N\"othnitzer Stra{\ss}e 38, 01187 Dresden, Germany}
\abstract{We consider a long-range model of coupled phase-only
oscillators subject to a local potential and evolving in presence of
thermal noise. The model is a non-trivial generalization of the
celebrated Kuramoto model of collective synchronization. We demonstrate by exact results and numerics a
surprisingly rich
long-time behavior, in which the system settles into either a stationary state that could be in or out of equilibrium and
supports either global synchrony or absence of it, or, in a
time-periodic synchronized state. The system shows both continuous and
discontinuous phase transitions, as well as an interesting reentrant transition in which the system 
successively loses and gains synchrony on steady increase of the
relevant tuning parameter.}
\date{\today}
\pacs{05.45.Xt}{Synchronization; coupled oscillators}
\pacs{05.70.Fh}{Phase transitions: general studies}
\pacs{05.70.Ln}{Nonequilibrium and irreversible thermodynamics}
\maketitle

\section{Introduction}
The issue of how self-sustained oscillators of
widely different natural frequencies when interacting {\it weakly} with
one another may {\it spontaneously} oscillate at a common frequency has intrigued
researchers since one of its first documented observations by the Dutch
physicist Christiaan Huygens in the 17th century. This phenomenon of collective synchrony is commonly observed in nature, e.g.,
in metabolic synchrony in yeast cell suspensions \cite{Bier:2000}, among
groups of fireflies that flash in unison \cite{Buck:1988}, in animal
flocking behavior \cite{Ha:2010},
and perhaps most importantly of all, in cardiac pacemaker cells whose
synchronized firings keep the heart beating and life going on \cite{Winfree:1980}.
Synchrony is much desired in man-made systems, e.g., in parallel computing in computer science, whereby
processors must coordinate to finish a task on time, and in electrical
power-grids constituting generators that must run in synchrony so as to
be tuned in frequency to that of the grid
\cite{Filatrella:2008,Dobson:2013}. However, synchrony could also be hazardous,
e.g., in neurons, leading to impaired brain function in Parkinson's
disease and epilepsy. For an overview, see Refs. \cite{Pikovsky:2001,Strogatz:2003}.

Over the years, the celebrated Kuramoto model has served as a paradigmatic model to study
analytically the emergence of synchrony in many-oscillator
systems \cite{Kuramoto:1984}; for a review, see Refs. \cite{Strogatz:2000,Acebron:2005,Gupta:2014}. The model comprises $N$ phase-only oscillators of
distributed natural frequencies that are globally coupled
through the sine of their phase differences. The $i$th
oscillator, $i=1,2,\ldots,N$, has its phase $\th_i \in [0,2\pi)$ evolving according to the equation 
\be
\frac{{\rm d}\theta_{i}}{{\rm
d}t}=\omega_i+\frac{K}{N}\sum_{j=1}^{N}\sin(\theta_{j}-\theta_{i}),
\label{eom-Kuramoto}
\ee
where $\omega_i$ is the natural frequency of the $i$th oscillator. Here,
$K >0$ is the coupling constant, and the second term on the right hand side (rhs) describes an
attractive all-to-all interaction between the oscillators, with the
factor $1/N$ making the model well behaved in the limit $N\to \infty$. 

For a
given frequency distribution $g(\omega)$ common to all, the
tendency of each oscillator to oscillate independently at its own
natural frequency is opposed by the global coupling among them.
Indeed, the latter favors equal phases for the oscillators, $\th_i=\th_j
~\forall~i \ne j$, thereby promoting global synchrony in which all the
oscillator phases turn around in time at a common frequency given by the mean $\Omega$
of the distribution $g(\omega)$. A measure of the amount of
synchrony in the system is given by the order parameter $r\equiv |\mathbf{r}|=
\sqrt{r_x^2+r_y^2}$, where the vector $\mathbf{r}$ is given by $\mathbf{r}= (r_x,r_y)\equiv
\sum_{j=1}^N (\cos \th_j,\sin \th_j)/N$ \cite{Kuramoto:1984}.

The Kuramoto model has been mostly studied for a
unimodal $g(\omega)$, i.e., one that is symmetric about the mean
$\Omega$, and which decreases monotonically to zero with increasing $|\omega -
\Omega|$. In this case, it has been shown that for a given $K$ and in the
limit $N \to \infty$, on tuning the width $\sigma$ of $g(\omega)$, the dynamics at long times favors a
synchronized phase ($r\ne0$) at low $\sigma$, and an unsynchronized phase ($r = 0$) at high $\sigma$, with a continuous transition between the two
phases at the critical value $\sigma_c\equiv\pi K g_1(\Omega)/2$
\cite{Strogatz:2000,Acebron:2005,Gupta:2014}, where $g_1$ is the
distribution obtained from $g$ by setting in the latter the width to
unity.

Several extensions of the Kuramoto model have been studied, including
the case with inertia for which the dynamics of phases 
is governed by second-order equations in time (see Ref. \cite{Gupta:2014} for a review). The main purpose
of these studies has been to investigate the robustness of
synchronization properties and change in their characteristics, as well
as in the thermodynamic phase diagram, in models that may reproduce realistic systems. A natural extension considered in this paper is represented by
the presence of a local potential acting as a torque on each oscillator.

\section{Model and main results} In this paper, we consider the case of identical oscillators,
$g(\omega)=\delta(\omega-\Omega)$, in the Kuramoto model, with an
additional thermal noise and a local potential
acting on the individual oscillators. The underlying equation of motion
is 
\be
\frac{{\rm d}\theta_{i}}{{\rm d}t}=\Omega+\frac{K}{N}\sum_{j=1}^{N}\sin(\theta_{j}-\theta_{i})-W\sin
2\th_i+\eta_{i}(t),
\label{eom}
\ee
where the third term on the rhs describes a torque due to a local
potential $-(W/2) \cos (2\th_i)$, with $W \ge 0$ quantifying its strength, while
$\eta_{i}(t)$ is a Gaussian, white noise, with
$\langle\eta_{i}(t)\rangle=0,~\langle\eta_{i}(t)\eta_{j}(t')\rangle=2T\delta_{ij}\delta(t-t')$.
The temperature $T$ is measured in units of the Boltzmann constant,
while angular brackets denote averaging over noise realizations.

The thermal noise may be interpreted as the effect of a coupling to the
external environment \cite{Gupta:2014}, but may also be seen as
accounting for stochastic fluctuations of the natural
frequencies in time \cite{Sakaguchi:1988}. The local potential alone,
which has its minima at $\th=0$ and $\th=\pi$, tends to ``pin" the $\th$'s to these
values, thus playing the role of a pinning potential, as was considered
in Ref. \cite{Strogatz:1989}. The latter work investigated the case of the
pinning potential $\sim \sin(\alpha_i-\theta_i)$ that tends to pin the phase of the $i$th
oscillator to the corresponding {\it random} pinning angle $\alpha_i$. With
every oscillator pinned to a random angle, such a pinning
potential evidently favors a static, {\it disordered} arrangement of the
phases. By contrast, the pinning potential in our case favors a static, {\it ordered} arrangement of the
phases.

We now remark on some known limits of the dynamics (\ref{eom}). For $W=T=0$, the dynamics reduces to that of the Kuramoto
model (\ref{eom-Kuramoto}), in the particular case of the latter in
which all the oscillators have the same
frequency. In this work, we consider $T>0$. For $\Omega=W=0$, Eq.
(\ref{eom}) describes the overdamped dynamics of the so-called
Brownian mean-field model \cite{Chavanis:2014}. The latter model
was introduced as a
generalization of the energy-conserving microcanonical dynamics of the Hamiltonian
mean-field (HMF) model \cite{Antoni:1995}, a paradigmatic system to study static and
dynamic properties of long-range interacting (LRI) systems\footnote{Long-range interacting (LRI) systems are those in which the
inter-particle potential decays slowly with the separation $r$ as
$r^{-\alpha}$ for large $r$, with $0 \le \alpha < d$ in $d$ spatial
dimensions \cite{Campa:2009,Bouchet:2010,Campa:2014}.}, to the
situation where the system exchanges energy by interacting with a heat bath. 

For $\Omega\ne0$ (respectively, $\Omega=0$), the dynamics (\ref{eom})
relaxes at long times to a non-equilibrium stationary state (NESS)
(respectively, equilibrium). Unlike equilibrium, NESSs do not
respect detailed balance and support fluxes of conserved
quantities in the system. Studying NESSs constitutes an active area of research in modern
statistical mechanics, where a challenge is to achieve a level of understanding akin to the
one established for equilibrium systems \cite{Privman}.
The local potential term in Eq. (\ref{eom}) does not allow to get rid of
the effect of the natural frequency by studying the dynamics in a frame rotating uniformly with frequency $\Omega$, thereby
recovering equilibrium dynamics, as is possible with $W=0$ \cite{Gupta:2014}.

The Kuramoto model (\ref{eom-Kuramoto}) subject to a periodic forcing $F \sin (\sigma t -\theta_i)$ has been studied
in Refs. \cite{Sakaguchi:1988,Ott:2008,Childs:2008} to address the issue of forced synchronization. The governing equation when
viewed in the frame rotating with the forcing frequency $\sigma$ reads
${\rm d}\theta_{i}/{\rm d}t=\omega_i-\sigma+(K/N)\sum_{j=1}^{N}\sin(\theta_{j}-\theta_{i})-F\sin \th_i$. Comparing with
Eq. (\ref{eom}), one immediately realizes the differences: in our model, all the oscillators have the same intrinsic frequency
$\Omega$, the parameter $\sigma=0$, there is thermal noise, and moreover, the additional term instead of being $-\sin \theta_i$
is $-\sin 2\theta_i$. The Kuramoto model (\ref{eom-Kuramoto}) with an additional external drive $f_e$ and a pinning potential
$-a \cos \theta_i+(b/2) \cos 2\theta_i$ on the rhs has been studied in Ref. \cite{Um:2012} to discuss how an ensemble of noiseless
globally-coupled pinned oscillators with distributed intrinsic frequencies is capable of rectifying spatial disorder, thus acting
as a rectifier \cite{Reimann:2002}, and exhibiting the phenomenon of absolute negative mobility \cite{Speer:2007}. In contrast
to this model, our model has all the oscillators with the same intrinsic
frequency (as we comment below, this implies that the
current has always the same sign as that of the external drive $\Omega$), and furthermore, it has stochastic noise in it. 

Before proceeding with an analysis of the model (\ref{eom}), we first set $K=1$ without loss of generality, and discuss the dynamical features and expected
behavior of the model at long times. Consider first the low-$T$ limit of the dynamics.
In the absence of the local potential (i.e., with $W=0$), the system goes to a trivially synchronized state\footnote{In the following, we use interchangeably the
terms ``state" and ``phase" to mean an ensemble of
microscopic configurations distributed with a given density.}.
Namely, it reduces to the Kuramoto model with identical frequency $\Omega$ of the oscillators.
Correspondingly, the order parameter $r$ is close to
unity, but the vector $\mathbf{r}$ varies {\it periodically in
time} with period $\Omega$. 
On the other hand, with $W>0$ and $\Omega=0$, the following occurs. As mentioned above, the local potential alone tends to make the
$\th$'s crowd around $\th=0,\pi$, while the global coupling alone tends to make
$\th_i=\th_j~\forall~i,j$. Thus, together, and in the absence of the
natural frequency term, the two terms lead to a static
arrangement of $\th$'s around either $0$ or $\pi$, resulting at long
times in a stationary state that is fully synchronized.
This tendency is opposed by the frequency term that tends to drive the phases at a 
constant angular velocity $\Omega$, thus leading, for sufficiently large $\Omega$, to a time-periodic
synchronized state. Increasing the temperature results in enhanced
thermal fluctuations that tend to destroy synchrony, be it in the
stationary or in the time-periodic state. The interplay between these
competing tendencies leads to a surprisingly rich phase
diagram, as we demonstrate below. 

Specifically, we show
that the system settles into either a stationary state that
could be in equilibrium or in a NESS and may or may not support global
synchrony, or, in a time-periodic synchronized state. In presence of
noise and other competing aspects of the dynamics, in the latter state 
the order parameter $r$ is not constant in time but shows time-periodic oscillations.
We also show that
the system exhibits both continuous and discontinuous transitions between 
different phases, as
well as an interesting reentrant transition in which the system
successively loses and gains synchrony on steady increase of the relevant tuning parameter.
It is remarkable that in the absence of a general framework to study NESSs akin to the one due to
Boltzmann and Gibbs for equilibrium states, our
tools and analysis allow to obtain {\it exact} analytical
results for the NESSs of the nontrivial dynamics (\ref{eom}). 

We remark that in terms of the $(r_x,r_y)$,
Eq. (\ref{eom}) becomes
\be
\frac{{\rm d}\theta_{i}}{{\rm d}t}=\Omega-r_x\sin \th_i+r_y \cos \th_i-W\sin
2\th_i+\eta_{i}(t),
\label{eom-r}
\ee
which makes it evident that the dynamics is effectively that of a single
oscillator evolving in a self-consistent mean field of strength $r$
produced by all the oscillators; the parameters characterizing the dynamics are $\Omega,T,W$.

\section{Analysis in the limit $N \to \infty$ and the stationary state}
In the limit $N\to\infty$,
the state of the system is characterized 
by the single-oscillator distribution $f(\theta,t)$, defined such
that $f(\theta,t){\rm d}\theta$ gives the probability at time $t$ that
a randomly chosen oscillator has its phase between $\theta$ and
$\theta+{\rm d}\theta$. One has $f(\theta+2\pi,t)=f(\theta,t)$, and the
normalization $\int_0^{2\pi}{\rm d}\theta\,f(\theta,t)=1~\forall~t$.
The time evolution of $f(\theta,t)$ is given
by the Fokker-Planck (FP) equation that may be straightforwardly written
down from Eq. (\ref{eom-r}):
\be
\frac{\partial f}{\partial
t}=-\frac{\partial[(\Omega-r_{x}\sin\theta+r_{y}\cos\theta-W\sin 2\theta)f]}{\partial\theta}+T\frac{\partial^{2}f}{\partial\theta^{2}},
\label{fpequation}
\ee
where $r_x,r_y$ are now functionals of $f(\th,t):~
(r_{x},r_{y})\equiv(r_x,r_y)[f]=\int
{\rm d}\theta\,(\cos\theta,\sin\theta)f(\theta,t)$.
When exists, the stationary solution of Eq.
(\ref{fpequation}) is obtained by setting the left hand
side (lhs) to zero; the solution
reads \cite{Sakaguchi:1988,Gupta:2014}
\be
f(\theta)=Be^{\frac{v(\theta)}{T}}\left[1+(e^{-2\pi\Omega/T}-1)\frac{\int_{0}^{\theta}{\rm
d}\theta'\,
 e^{-\frac{v(\theta')}{T}}}{\int_{0}^{2\pi}{\rm d}\theta'\, e^{-\frac{v(\theta')}{T}}}\right],
\label{fpequationsteady}
\ee
with $v(\th)\equiv
v[f](\theta)=\Omega\theta+r_{x}\cos\theta+r_{y}\sin\theta+(W/2)\cos
2\theta$, 
while the constant $B$ is fixed by the normalization of $f(\theta)$.
The stationary quantities $r_{x,y}$ are determined self-consistently:
\be
(r_{x},r_{y}) \equiv \int
{\rm d}\theta\,(\cos\theta,\sin\theta)f(\theta).
\label{selfcons}
\ee

The stationary unsynchronized solution $f_0(\th)$ with $r_x=r_y=0$ always
exists, but it may be unstable in regions of the parameter space
$(\Omega,T,W)$; Eq. (\ref{fpequationsteady}) yields
\bea
\label{fpequationsteadyr0}
\hspace{-1cm}&&f_0(\theta)=Be^{\frac{v_0(\theta)}{T}}\left[1+(e^{-2\pi\Omega/T}-1)\frac{\int_{0}^{\theta}{\rm
d}\theta'\,
 e^{-\frac{v_0(\theta')}{T}}}{\int_{0}^{2\pi}{\rm d}\theta'\,
 e^{-\frac{v_0(\theta')}{T}}}\right],
\eea
with $v_0(\th)\equiv v_0[f](\theta)=\Omega\theta+(W/2)\cos 2\theta$.
For $W=0$, one has the homogeneous
distribution, $f_0(\th)=1/(2\pi)$, but for $W>0$, although
unsynchronized, $f_0(\theta)$ is {\it not} homogeneous in $\th$.
As mentioned earlier, one has an equilibrium stationary state for
$\Omega=0$ (i.e., on the $(T,W)$-plane), and a NESS for $\Omega \ne
0$.

One may define a current
$J\equiv\langle\dot{\theta}\rangle/(2\pi)$ \cite{Um:2012}, to characterize the motion of
the oscillator phases. Summing Eq. (\ref{eom-r}) over $i$ and averaging over noise gives
$J(t)=[\Omega-W\langle\sin2\theta\rangle(t)]/(2\pi)$. In the stationary state, the stationary current
$J_{\rm st}$ is obtained from Eq. (\ref{fpequationsteady}) as $J_{{\rm
st}}=BT[1-e^{-2\pi\Omega/T}]/\int_{0}^{2\pi}d\theta'\thinspace e^{-v(\theta')/T}$,
which implies zero current for $\Omega=0$; this is expected since in
this case, the system is in equilibrium. Also, for $\Omega \ne 0$, when
the system is in a NESS, we have that $J_{\rm st}$ has the same sign as
that of $\Omega$.
Note that these features hold irrespective of whether the stationary state is synchronized or unsynchronized. In the time-periodic state,
the current is obtained either from the equations of motion (\ref{eom-r}) or by using the Fokker-Planck equation (\ref{fpequation}).

\section{Linear stability of the stationary unsynchronized state} Let us now study
the linear stability of the stationary unsynchronized state $f_0(\theta)$. To
this end, we linearize Eq.
(\ref{fpequation}) around $f_0(\th)$, by writing $f(\theta,t)=f_0(\theta)
+\delta f(\theta,t);~|\delta f| \ll 1$, and keeping the zeroth and
first-order terms in 
$\delta f$. The zeroth-order terms cancel; the first-order
terms give
\bea
&&\frac{\partial \delta f}{\partial t}
=-\frac{\partial}{\partial \theta}\left[(\Omega-W\sin 2\theta)\delta
f\right]\nonumber \\
&&+\frac{\partial}{\partial
\theta}\left[(r_x\sin\theta-r_y\cos\theta)f_0(\th)\right]
+T\frac{\partial^2 \delta f}{\partial\theta^2},
\label{fpequationlinear}
\eea
with $(r_x,r_y)=\int \dd
\theta\,(\cos\theta,\sin\theta)\delta f(\theta,t)$.
Now, $f_0(\theta)$ is stable if the eigenvalue spectrum of the linear
operator $\mathcal{L}$ defined by the rhs of Eq.
(\ref{fpequationlinear}) has a negative
real part. 

To study the linear problem, we compute the matrix elements
$\mathcal{L}_{k',k}$ of
$\mathcal{L}$ with respect to the set of complete basis
constituted by the functions $\exp (ik\theta)$, with $k$ an integer, for 
functions in $[0,2\pi)$. Using Eq. (\ref{fpequationlinear}), we have 
\bea
\mathcal{L}_{k',k}\hspace{-0.5cm}&&\equiv\frac{1}{2\pi} \int_0^{2\pi} \dd \theta \,
e^{-ik'\theta} \mathcal{L} e^{ik\theta} \nonumber \\
\hspace{-1cm}&&=-(ik\Omega+k^2T)\delta_{k',k} +\frac{ik'W}{2\pi}\int \dd \theta \, \sin 2\theta e^{i(k-k')\theta}
\nonumber \\
\hspace{-1cm}&&+\frac{ik'}{2\pi}\int \dd \theta \, \left[r_x \sin \theta - r_y \cos \theta\right]f_0(\theta) e^{-ik'\theta},
\label{matrelem_b}
\eea
with $(r_x,r_y) = \int \dd \theta\,(\cos\theta,\sin\theta)e^{ik\theta}$.
Using the relations $\int \dd \theta \, \cos \theta e^{ik\theta} = \pi
(\delta_{k,1}+\delta_{k,-1}),
\int \dd \theta \, \sin \theta e^{ik\theta} = i\pi
(\delta_{k,1}-\delta_{k,-1}),\int \dd \theta \, \sin 2\theta e^{ik\theta}
= i\pi (\delta_{k,2}-\delta_{k,-2})$ in Eq. (\ref{matrelem_b}) gives
\bea
&&\mathcal{L}_{k',k}=-(ik\Omega +k^2T)\delta_{k',k} +\frac{W}{2}k' (\delta_{k',k+2}-\delta_{k',k-2})
\nonumber \\
&&+\frac{ik'}{2}\int \dd \theta \,
\Big[(\delta_{k,1}+\delta_{k,-1}) \sin \theta \nonumber \\
&&- i(\delta_{k,1}-\delta_{k,-1}) \cos \theta\Big]f_0(\theta) e^{-ik'\theta}.
\label{matrelem_c}
\eea
Using $\sin \theta \pm i\cos \theta = \pm i e^{\mp i\theta}$, and
introducing the Fourier components of $f_0(\theta)$, $\widetilde{f}_p \equiv
\int_0^{2\pi} \dd \theta \, e^{-ip\theta}f_0(\theta)$, 
we arrive at the following final expression of the matrix elements:
\bea
&&\mathcal{L}_{k',k}=-(ik\Omega +k^2T)\delta_{k',k} +\frac{W}{2}k' (\delta_{k',k+2}-\delta_{k',k-2})
\nonumber \\
&&+\frac{k'}{2} \left[\delta_{k,1}\widetilde{f}_{k'-1}
-\delta_{k,-1}\widetilde{f}_{k'+1}\right].
\label{matrelem_fin}
\eea
For $W=0$, we have $f_0(\theta)=1/(2\pi)$, giving
$\widetilde{f}_p=\delta_{p,0}$, so that the matrix becomes diagonal,
\be
\mathcal{L}_{k',k} = \left[ -ik\Omega -k^2T +\frac{1}{2} (\delta_{k,1}+\delta_{k,-1})\right]\delta_{k',k},
\label{matrelem_fin_w0}
\ee
which readily gives the eigenvalues. Before proceeding, let us adopt the following order for the rows
and columns of the matrix. The indices $k,k'$ take integer values; we order the indices according
to $k,k'=0,1,-1,2,-2,3,-3,\dots$. With this ordering, we see that all elements of the first row of the
matrix vanish, i.e., $\mathcal{L}_{0,k}=0~\forall~k$. In the first
column, only two elements do not vanish: $\mathcal{L}_{2,0}=\mathcal{L}_{-2,0}=W$.

Now, the order of the matrix $\mathcal{L}$ being infinite, we consider
in the following the matrix of a finite size obtained by truncating
the infinite one at a given maximum value $k=k_{\rm max}$; the order of this finite
matrix is $2k_{\rm max}+1$. We study the eigenvalues of this truncated
matrix; we have chosen a value for $k_{\rm max}$ that is large enough
that our results, as far as the determination of stability is concerned, do not change for larger $k_{\rm max}$. Since the elements in
the first row of the matrix vanish, there is always a zero eigenvalue; however, the stability of $f_0(\theta)$
is determined by the other eigenvalues, i.e., $f_0(\th)$ is stable if the real part of all the other eigenvalues is negative,
while it is unstable if there is at least one eigenvalue with a positive real part.

To see that the zero eigenvalue may be disregarded, we reason as follows. Since the original FP equation conserves the
normalization of $f$, and $f_0(\th)$ is normalized, we have to study the linearized equation (\ref{fpequationlinear}) for functions $\delta f(\theta,t)$
that satisfy $\int_0^{2\pi} \dd \th \, \delta f(\th,t)=0 ~\forall~t$.
Now, the eigenfunction
$\delta f$ corresponding to the zero eigenvalue is obtained directly
from Eq. (\ref{fpequationlinear}) by setting the lhs to zero. From the definition of $f_0(\theta)$, we see that this linear equation is solved by
$\delta f(\theta) \propto f_0(\theta)$. Since $f_0(\theta)$ is positive
definite, such a $\delta f$ has a non-vanishing
integral over $[0,2\pi)$ (thus, it has a nonzero Fourier component for
$k=0$), and is therefore not suitable for our stability problem. On the
other hand, each eigenfunction $\delta f$ corresponding to a non-vanishing eigenvalue
must have a zero Fourier component for $k=0$: in fact, $\sum_k
\mathcal{L}_{0,k} a_k=0$ for any vector $a_k$, as we have seen,
then this cannot be equal to $\lambda a_0$ if both $\lambda$ and $a_0$ are different from $0$. In conclusion, any
non-vanishing eigenvalue of the matrix $\mathcal{L}$ has a corresponding eigenfunction whose first component (i.e., that
corresponding to $k=0$), is $0$, and thus, it belongs to the class suitable for our stability problem. This also implies,
as a byproduct, that the matrix $\mathcal{L}$ and the matrix of order
$2k_{\rm max}$ obtained by deleting the first row and the first
column of $\mathcal{L}$ have the same non-vanishing eigenvalues.

Summarizing, $f_0(\th)$ is stable if all the non-vanishing eigenvalues
of $\mathcal{L}$ have a negative real part. For our computations,
for any chosen values of $(T,W)$, we vary $\Omega$, and compute
numerically the eigenvalues of $\mathcal{L}$, thus determining the
parameter regimes where $f_0$ is stable. When $f_0$ is unstable, the
system may settle into a synchronized state that is either stationary or
time-periodic.

\section{Linear stability of the stationary synchronized state}
We now discuss how to obtain the linear stability threshold of the synchronized state (\ref{fpequationsteady}). This threshold
should be computed by linearizing the Fokker-Planck equation
(\ref{fpequation}) about the state (\ref{fpequationsteady}), along the
lines followed above to compute the stability threshold of the
unsynchronized state. Given the complexity of the state (\ref{fpequationsteady}), such a task
turns out to be quite formidable, and, hence, we resort to a different
strategy, as detailed below. 

To compute the threshold for given
values of $T$ and $W$, we vary $\Omega$, and obtain numerically the
corresponding non-zero values of $r_x$ and $r_y$ that satisfy the self-consistent
equations (\ref{selfcons}). Operatively, this is
done as follows. For $\Omega=0$, the
corresponding non-zero $r_x$ is easily found numerically, while one has $r_y=0$.
Then, for $\Omega>0$, we increase $\Omega$ in small steps of size $\dd
\Omega$, and compute correspondingly the new values of $r_x$ and $r_y$ after every step by using
the partial derivatives of $f(\theta;T,W,\Omega,r_x,r_y)$ with respect
to the parameters. Here, we have written explicitly the parameters on
which the solution (\ref{fpequationsteady}) depends, as $f(\theta;T,W,\Omega,r_x,r_y)$ (the normalizing factor $B$ then
also depends on the parameters). Having reached the $\Omega$ value corresponding to the
stability threshold of the synchronized state is signaled by the fact
that the above procedure does not find any more non-zero $r_x$ and $r_y$.

\begin{figure}[!h]
\centering
\includegraphics[width=7.5cm]{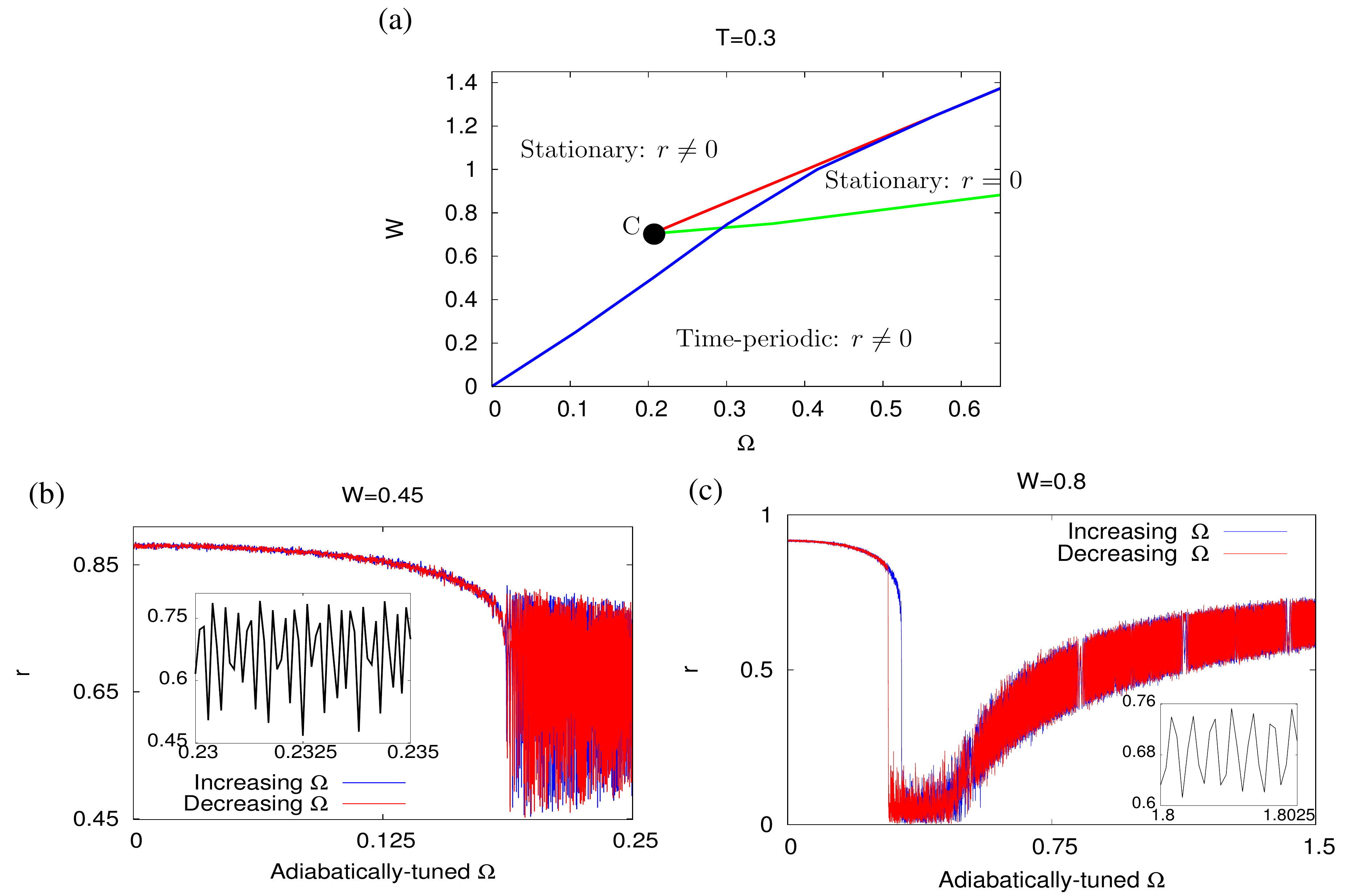}
\caption{(a) Phase diagram in the $(\Omega,W)$-plane for $T=0.3$. Here, the red line (respectively, the green line)
corresponds to loss of stability of the stationary unsynchronized 
state and transition to the stationary synchronized state (respectively,
the time-periodic state) at smaller (respectively, larger) $\Omega$. The
blue line refers to the stability threshold of the stationary
synchronized state.
(b),(c): Numerical simulation results for $r$ vs. adiabatically-tuned $\Omega$. The system
size is $N=10^4$. Panel (b) corresponds to a value of
$W$ lower than the one corresponding to the point $C$ in (a). Panel (c)
refers to a value of $W$ larger than the
one corresponding to the point $C$, and specifically, to a value of $W$
at which on increasing $\Omega$, the green line in (a) is met after the blue line. The insets in panels (b) and (c)
demonstrate time-periodic behavior of $r$ in the relevant $\Omega$-regime. } 
\label{fig:T0-3}
\end{figure}

Figure \ref{fig:T0-3}(a) shows representative results of our analytic
computation, in which (i) the red line indicates loss of stability of
the stationary unsynchronized state $f_0(\th)$ and transition to the 
stationary synchronized state at smaller $\Omega$, (ii) the blue line indicates loss of stability of
the stationary synchronized state (\ref{fpequationsteady}), and
consequently, a transition at larger $\Omega$ either to the stationary 
unsynchronized state or to the time-periodic state (see below the
different cases), and (iii) the green line indicates loss of stability of
the stationary unsynchronized state $f_0(\th)$ and transition to the 
time-periodic state at larger $\Omega$. Our computation indicates that the red and green lines
emerge from a common point $C=C(\Omega,W,T)$, while the blue line starts at the origin ($\Omega=W=0$),
extending in a region where the red and green lines do not exist. In this latter region, the
blue line corresponds to a transition between the stationary synchronized and the time-periodic state. This transition
may not be called a phase transition since the latter is usually meant to
refer to a transition between two states that are both stationary,
besides having different macroscopic properties. The red and blue lines eventually
merge with one another, see Fig. \ref{fig:T0-3}(a). The point $C$ approaches the origin
($\Omega=W=0$) as $T$ approaches $1/2$ from below;
concomitantly, the green line approaches the $\Omega$-axis, while the
red line becomes more inclined towards the $W$-axis, approaching the blue line and disappearing for
$T >1/2$; this suggests that for $T \ge 1/2$, the late-time
dynamics does not support a time-periodic state. 

We now report on numerical simulations of the dynamics\footnote{All numerical simulations reported in this paper
involved employing a first-order Euler algorithm to integrate the
equation of motion (\ref{eom}), with timestep ${\rm d}t=0.01$.}. At
fixed values of $(W,T)$, we perform two sets of simulations. In one set, we start at $\Omega=0$ from a
state with $r=1$, prepared by setting $\th_i=0~\forall~i$, and then tune
$\Omega$ adiabatically to a large-enough value, while simultaneously monitoring $r$ as a
function of $\Omega$. In the other set, we start with the state obtained
as the final state of the first set of simulations, and then
adiabatically decrease $\Omega$ to zero. 
Adiabatic tuning of $\Omega$ ensures that the system has sufficient time
to attain stationarity before the value of $\Omega$ changes
significantly. The results of simulations are displayed in Fig.
\ref{fig:T0-3}, panels (b) and (c) at a temperature below $1/2$, namely,
$T=0.3$. We now discuss in detail the results presented in these panels.

Figure \ref{fig:T0-3}(b) corresponds to a value of
$W$ lower than the one corresponding to the point $C$; we note that
consistent with the phase diagram in (a), the system goes over from the 
stationary synchronized to the time-periodic state by crossing the blue
line in (a). 

Figure \ref{fig:T0-3}(c) refers to a value of $W$ larger than the
one corresponding to the point $C$, and specifically, to a value at
which on increasing $\Omega$, the green line in (a) is met after the
blue line. The results show that consistent with our computation, on increasing $\Omega$, the system goes
over from the stationary synchronized to the stationary unsynchronized state
(corresponding to crossing of the blue line in (a)), and
then goes over to the time-periodic synchronized state (corresponding to
crossing of the green line in (a)). Subsequently, on decreasing
$\Omega$, the system goes over to the stationary unsynchronized state by
crossing the green line in (a) and then to the stationary synchronized state by
crossing the red line in (a). This behavior suggests
that while the transition between the stationary synchronized and the
stationary 
unsynchronized state is discontinuous (as evidenced by the presence of
a hysteresis loop in (c)), the one between the stationary unsynchronized
and the time-periodic state is continuous (implied by the
absence of any hysteresis loop in (c)). We
see from Fig. \ref{fig:T0-3}(c) the
emergence of a feature induced by the local potential, which is absent in the Kuramoto model, namely, that of a 
reentrant transition in which the system successively loses and gains
synchrony on steady increase of $\Omega$. Similar reentrant transitions
between states that are stationary (unlike our case where one state is
time-periodic) are observed in the version of
the Kuramoto model that includes a phase lag in the
coupling and inertial effects \cite{Omelchenko:2012,Komarov:2014}.

Let us now discuss the situation for a value of $W$ larger than, but close to, the
one corresponding to the point $C$, and specifically, to a value at
which on increasing $\Omega$, the green line in Fig. \ref{fig:T0-3}(a) is met before the
blue line. Here, on increasing $\Omega$, the system goes over from the stationary synchronized
to the time-periodic state on crossing the blue line in (a) (note that
no change in behavior is expected on crossing the green line in (a)
while starting from the state $r=1$ and then adiabatically increasing
$\Omega$.) On decreasing $\Omega$, the time-periodic state goes over to
the stationary unsynchronized state on crossing the green line in (a)
and then to the stationary synchronized state on crossing the red line
in (a).

\begin{figure}[!h]
\centering
\includegraphics[width=7.5cm]{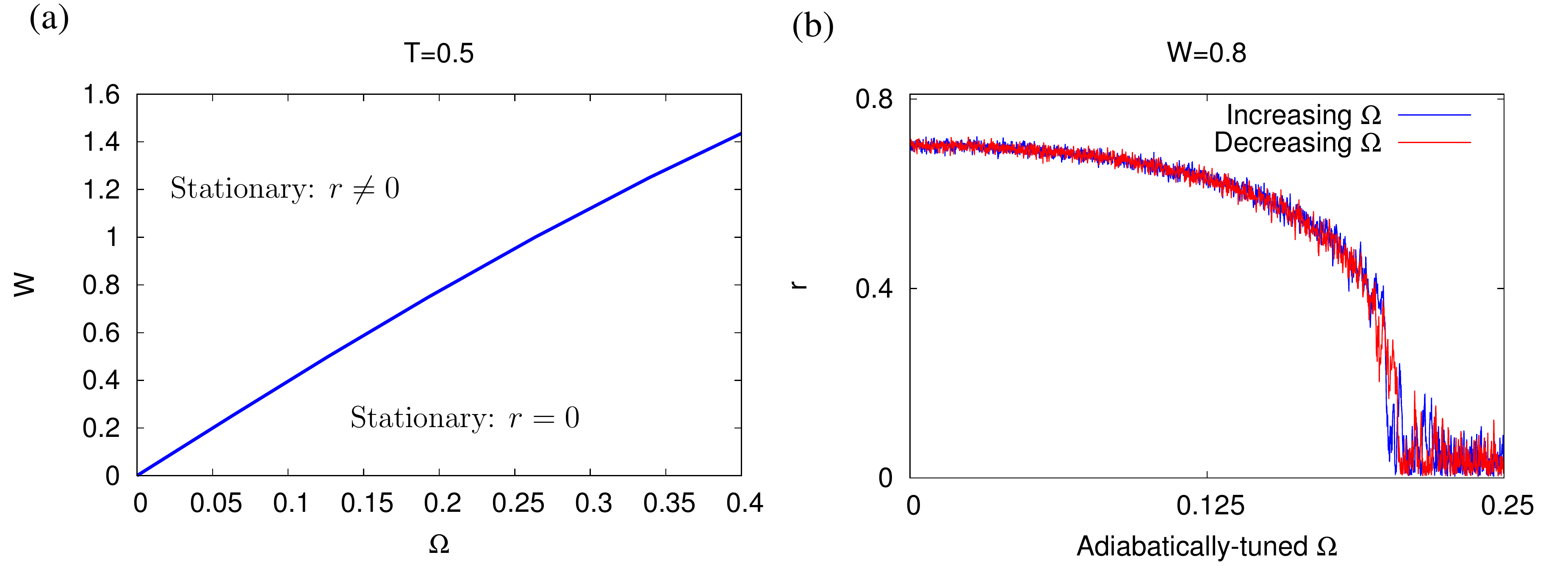}
\caption{(a) Phase diagram in the $(\Omega,W)$-plane for $T=0.5$. The blue line refers to the line of continuous
transition between the stationary synchronized and unsynchronized
states. (b) Numerical simulation results for $r$ as a function of
adiabatically-tuned $\Omega$ at a representative $W$. The system
size is $N=10^4$. }
\label{fig:T0-5}
\end{figure}

\begin{figure}[!h]
\centering
\includegraphics[width=7.5cm]{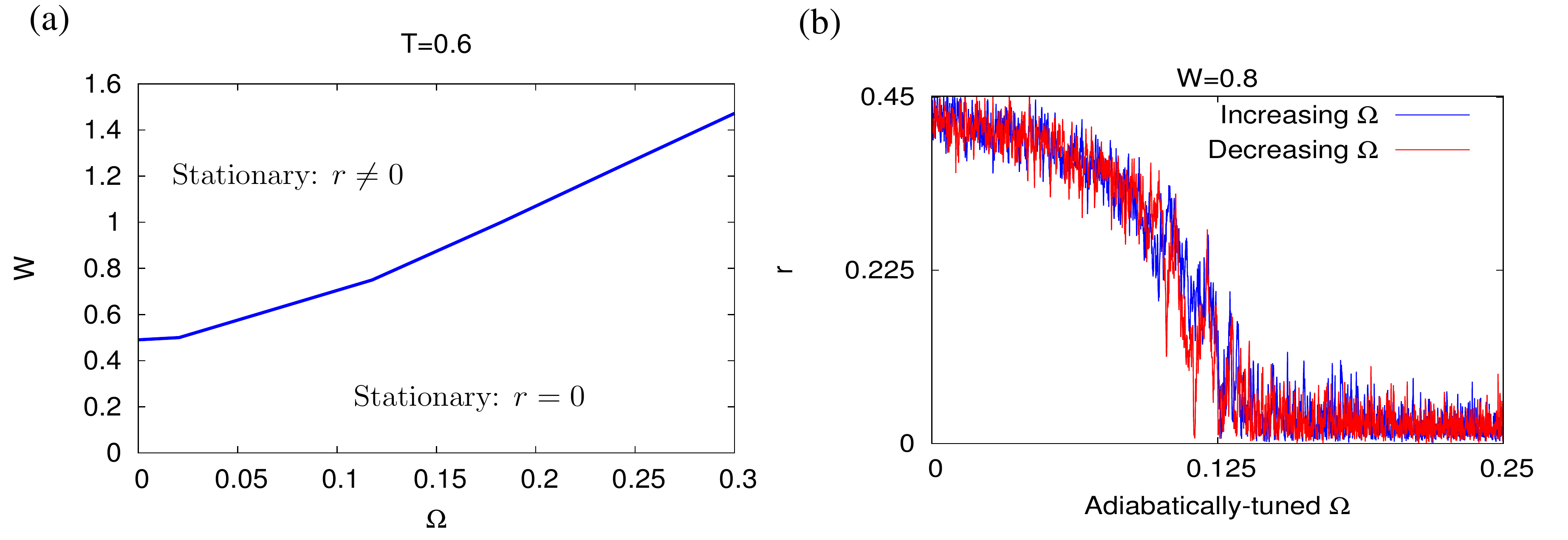}
\caption{(a) Phase diagram in the $(\Omega,W)$-plane for $T=0.6$. The blue line refers to a continuous
transition between the stationary synchronized and unsynchronized
states. (b) Numerical simulation results for $r$ as a function of
adiabatically-tuned $\Omega$ at a representative value of $W$ above the minimum of the
blue line in (a) (that occurs at $\Omega=0$); for $W$ values below this minimum, $r$ is equal to
zero at all $\Omega$. The system
size is $N=10^4$. }
\label{fig:T0-6}
\end{figure}

Till now we have discussed the behavior of our model for $T <1/2$. For
$T \ge 1/2$, our analysis presented below shows that one has a
continuous transition between the stationary synchronized and the
stationary unsynchronized state (we have already discussed that the time-periodic state disappears
as $T$ approaches $1/2$ from below).
\section{Continuous transition points between the stationary synchronized and
unsynchronized states}
For the determination of points of continuous transition where the
stationary unsynchronized state (\ref{fpequationsteadyr0}) becomes
synchronized, we adopt the usual strategy of expanding to first order the
self-consistent equation of the order parameter. The two
self-consistent equations (\ref{selfcons}) are rewritten as
\be
(r_x,r_y)=\int \dd \theta \, (\cos \theta, \sin \theta)
~f(\theta;T,W,\Omega,r_x,r_y).
\label{selfconsexp}
\ee
We now make an expansion of the function $f$ around $r_x=r_y=0$. Denoting with a subscript the partial derivatives
of $f$ with respect to the parameters, we then write
\bea
\hspace{-1.1cm}&&f(\theta;T,W,\Omega,r_x,r_y)=f(\theta;T,W,\Omega,0,0) \\
\hspace{-1.1cm}&&+ f_{r_x}(\theta;T,W,\Omega,0,0)r_x
+f_{r_y}(\theta;T,W,\Omega,0,0)r_y +O(r^2). \nonumber
\label{expanf}
\eea
Since the integral of either $\cos \theta$ or $\sin \theta$ multiplied by $f(\theta;T,W,\Omega,0,0)$ vanishes,
plugging the last expansion in Eq. (\ref{selfconsexp}), and defining $C_{xx} \equiv \int \dd \theta \, \cos \theta~f_{r_x}(\theta)$,
$C_{xy}\equiv \int \dd \theta \, \cos \theta~f_{r_y}(\theta), C_{yx}\equiv \int \dd \theta \, \sin \theta~f_{r_x}(\theta),
C_{yy}\equiv\int \dd \theta \, \sin \theta~f_{r_y}(\theta)$, (the parameters $(T,W,\Omega,0,0)$ in $f_{r_x}$ and $f_{r_y}$
are not displayed for visual clarity) the self-consistent equations may be written as
\be
C_{xx}r_x+C_{xy}r_y =r_x, ~C_{yx}r_x+C_{yy}r_y =r_y.
\label{selfconsmatc}
\ee
This system of equations has nontrivial solutions when the matrix $C$ has an eigenvalue equal to $1$, i.e.,
when $\det (C-I)=0$, with $I$ the identity matrix. The latter condition
gives an equation in $(\Omega,T,W)$ the
solution of which gives the surface of continuous transition in the
parameter space. The line in the $(T,W)$-plane given by the intersection
of this surface is the line of critical points between equilibrium states (since $\Omega=0$). In this case, the solution (\ref{fpequationsteady})
becomes $f(\theta;T,W,0,r_x,0))=B\exp[(r_{x}\cos\theta+(W/2)\cos 2\theta)/T]$, 
since $r_y=0$, and the points of continuous transition are given by $C_{xx}=1$.

The results of our computation are shown by blue lines in panel (a) of Figs.
\ref{fig:T0-5} and \ref{fig:T0-6} for $T=0.5$ and $T=0.6$, respectively.
Numerical simulation results for $r$ as a function of
adiabatically-tuned $\Omega$, displayed in the other panels of these
figures, show the absence of a hysteresis loop characteristic of a
discontinuous transition, but rather a variation that coincides, up to
fluctuations due to finite $N$ and thermal noise, for increasing and decreasing $\Omega$.
This observation supports our claim that the system undergoes a continuous transition
between the stationary synchronized and the stationary unsynchronized state on crossing the
blue line in (a) of Figs.
\ref{fig:T0-5} and \ref{fig:T0-6}.

We note that for our model, the current $J$ is always positive for $\Omega > 0$.
In the stationary state, $J_{{\rm st}}$ increases with $\Omega$ independently from the
synchronization property of the state. In the time-periodic states, $J$ is obviously periodic, with its average over a period increasing with $\Omega$.

\begin{figure}[!h]
\includegraphics[width=6.3cm]{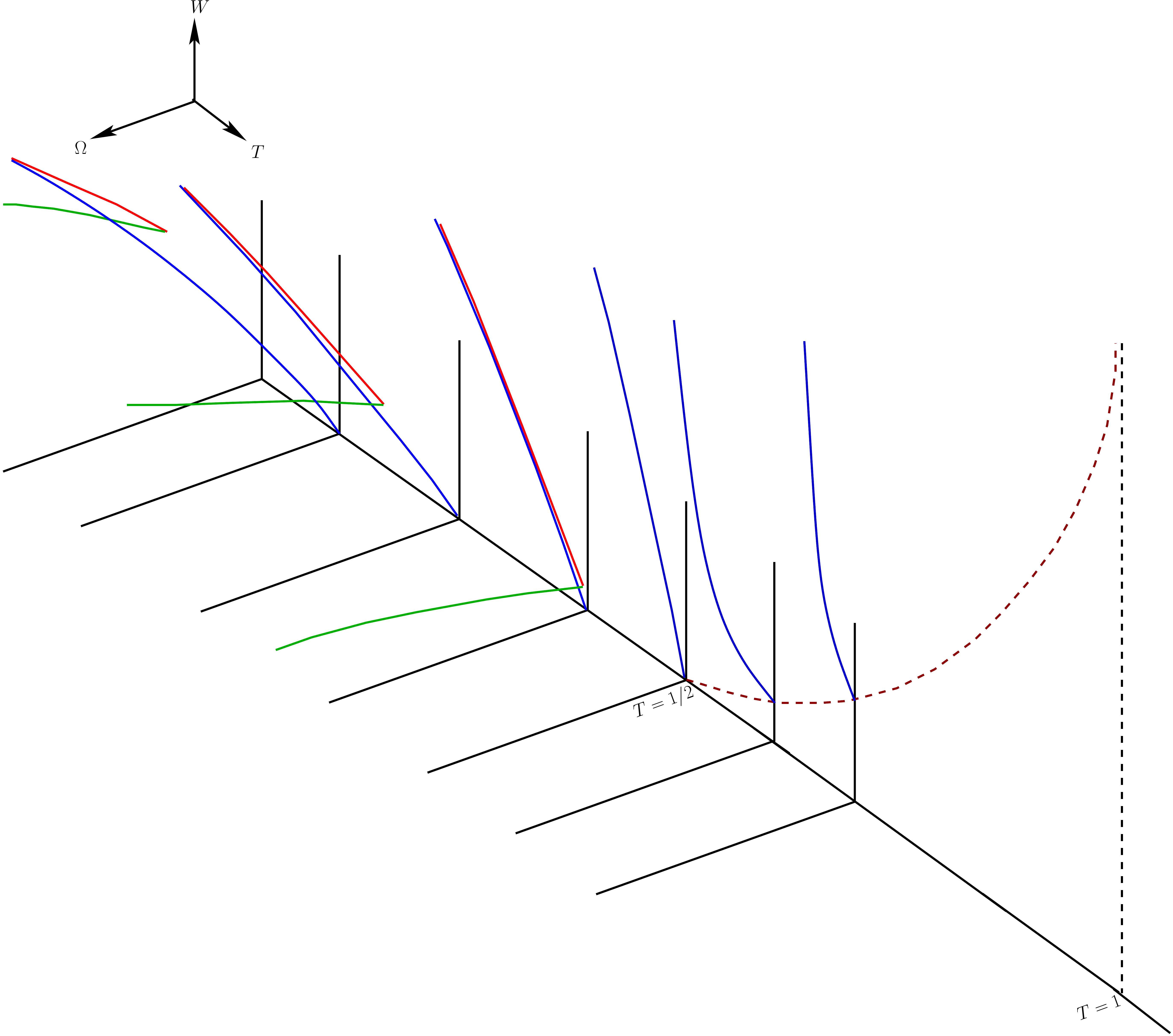}
\caption{Schematic phase diagram in the
$(\Omega,T,W)$ space, specifically, on a set of $(\Omega,W)$-planes for
increasing values of $T$.}
\label{fig:3d-ph-diag}
\end{figure}

Based on our results, we show in Fig. \ref{fig:3d-ph-diag} a schematic phase
diagram of our model in the $(\Omega,T,W)$ space. Note the point of continuous
transition for $\Omega=W=0$ given by $T=T_c=1/2$, which coincides with
the phase transition point for the BMF model to which the dynamics
(\ref{eom}) reduces under these limiting values of $\Omega$ and $W$.
With respect to Fig. \ref{fig:3d-ph-diag}, one may wonder as to how for
$T\ge 1/2$ does the point of intersection of the blue line of continuous
transition with the $(W,T)$-plane (i.e., $\Omega=0$) behave as a
function of $T$. We now show that this point, which starts at $(T=1/2,W=0)$, has the limiting
value $(T=1,W\to \infty)$, see the brown dotted line on the
$(T,W)$-plane in Fig. \ref{fig:3d-ph-diag}. In the limit $W \to \infty$, the oscillator phases have only 
two allowed values,
$\theta = 0$ and $\theta = \pi$, thus, $\cos \theta$ can be either $+1$ or $-1$. 
This implies that $r_y=0$, while $r=r_x$ may be easily computed by noting that for $\Omega=0, W\to \infty$, the
dynamics has the Gibbs-Boltzmann equilibrium stationary state:
$f(\theta=0)\propto \exp(r_x/T)$ and $f(\theta=\pi)\propto \exp(-r_x/T)$. We thus have
$r=\tanh(r/T)$. This transcendental equation
has a non-zero solution for $r$ provided $T \le 1$, giving the critical
temperature of transition as unity.
\section{Conclusions and perspectives}
In this paper, we studied a model system of globally-coupled oscillators
with an additional local potential acting on the individual oscillators,
and evolving in presence of thermal noise. We demonstrated by exact results and numerics a
surprisingly rich
long-time behavior, in which the system settles into either a stationary state that could be in or out of equilibrium and
supports either global synchrony or absence of it, or, in a
time-periodic synchronized state.
We note that without noise ($T=0$), the only possibility at long times would be given by
all oscillators at the same position, either in a stationary state ($\Omega < W$) or in
a time-periodic state ($\Omega > W$), thus a much less rich behavior.
While we have an exact expression for
the single-oscillator distribution in the stationary synchronized state
(Eq. (\ref{fpequationsteady})) and in the stationary unsynchronized
state (Eq. (\ref{fpequationsteadyr0})), both obtained by solving the
Fokker-Planck equation (\ref{fpequation}) in the stationary state, it is of interest to
obtain the distribution in the time-periodic state by solving the full
time-dependent Fokker-Planck equation.
It would be also interesting to
consider in place of the first-order dynamics (\ref{eom}) the
corresponding second-order dynamics that accounts for the finite
moment-of-inertia of the oscillators. In case of the Kuramoto model, the
corresponding second-order dynamics leads to rather drastic consequences
\cite{Gupta:2014-1,Olmi:2014,Olmi:2016,Campa:2015}, and an issue would be to investigate how
much of these are retained or modified by the presence of the local potential.
Another interesting direction to pursue would be to consider the
dynamics (\ref{eom}) with a non-mean-field coupling, i.e., with a coupling
that decays as a power-law of the separation between oscillators
occupying the sites of say a one-dimensional lattice
\cite{Gupta:2012}.

{\bf Acknowledgements:} We acknowledge fruitful discussions with S.
Ruffo and A. Torcini. SG acknowledges financial support of INFN Roma and
hospitality of Universit\`{a} di Roma ``La Sapienza".



\begin{thebibliography}{99}

\bibitem{Bier:2000}M. Bier, B. M. Bakker, and H. V. Westerhoff, Biophys.
J. {\bf 78},
1087 (2000).


\bibitem{Buck:1988}J. Buck, Quart. Rev. Biol. {\bf 63}, 265 (1988).

\bibitem{Ha:2010}S. Y. Ha, E. Jeong, and M. J. Kang, Nonlinearity {\bf
23}, 3139 (2010).

\bibitem{Winfree:1980}A. T. Winfree, {\it The Geometry of Biological
Time} (Springer, New York, 1980).

\bibitem{Filatrella:2008}G. Filatrella, A. H. Nielsen, and N. F.
Pedersen, Eur. Phys. J B {\bf 61}, 485 (2008).

\bibitem{Dobson:2013}I. Dobson, Nat. Phys. {\bf 9}, 133 (2013).

\bibitem{Pikovsky:2001}A. Pikovsky, M. Rosenblum, and J. Kurths, {\it
Synchronization: A Universal Concept in Nonlinear Sciences} (Cambridge
University Press, Cambridge, 2001).

\bibitem{Strogatz:2003}S. H. Strogatz,
{\it Sync} (Hyperion, New York, 2003).

\bibitem{Kuramoto:1984}Y. Kuramoto, \textit{Chemical oscillations, Waves
and Turbulence} (Springer, Berlin, 1984).

\bibitem{Strogatz:2000}S. H. Strogatz, Physica D {\bf 143}, 1 (2000).

\bibitem{Acebron:2005}J. A. Acebr\'{o}n, L. L. Bonilla, C. J. P.
Vicente, F. Ritort, and R. Spigler, Rev. Mod. Phys. {\bf 77}, 137
(2005).

\bibitem{Gupta:2014}S. Gupta, A. Campa, and S. Ruffo, J. Stat. Mech.: Theory Exp. R08001 (2014).

\bibitem{Sakaguchi:1988}H. Sakaguchi, Prog. Theor. Phys. {\bf 79}, 39
(1988).

\bibitem{Strogatz:1989}S. H. Strogatz, C. M. Marcus, R. M. Westervelt, and R. E. Mirollo, Physica D {\bf 36}, 23 (1989).


\bibitem{Chavanis:2014}P. H. Chavanis, Eur. Phys. J. B {\bf 87}, 120
(2014).

\bibitem{Antoni:1995}M. Antoni and S. Ruffo, Phys. Rev. E {\bf 52}, 2361 (1995).

\bibitem{Campa:2009}A. Campa, T. Dauxois, and S. Ruffo, Phys. Rep. {\bf
480}, 57 (2009).

\bibitem{Bouchet:2010}F. Bouchet, S. Gupta, and D. Mukamel, Physica A {\bf 389}, 4389 (2010).

\bibitem{Campa:2014}A. Campa, T. Dauxois, D. Fanelli, and S. Ruffo, {\it
Physics of Long-range Interacting Systems} (Oxford University Press,
Oxford, 2014).

\bibitem{Privman}V. Privman, {\it Nonequilibrium Statistical Mechanics
in One Dimension} (Cambridge University Press, Cambridge, 2005).

\bibitem{Ott:2008}E. Ott and T. M. Antonsen, Chaos {\bf 18}, 037113
(2008).

\bibitem{Childs:2008}L. M. Childs and S. H. Strogatz, Chaos {\bf 18}, 043128 (2008).


\bibitem{Um:2012}J. Um, H. Hong, F. Marchesoni, and H. Park, Phys. Rev. Lett. {\bf 108},
060601 (2012).

\bibitem{Reimann:2002}P. Reimann, Phys. Rep. {\bf 361}, 57 (2002).

\bibitem{Speer:2007}D. Speer, R. Eichhorn, and P. Reimann, Europhys. Lett. {\bf 79}, 10005
(2007).

\bibitem{Omelchenko:2012}O. E. Omel'chenko and M. Wolfrum, Phys. Rev.
Lett. {\bf 109}, 164101 (2012).

\bibitem{Komarov:2014}M. Komarov, S. Gupta, and A. Pikovsky, EPL {\bf
106}, 40003 (2014). 

\bibitem{Gupta:2014-1}S. Gupta, A. Campa, and S. Ruffo, Phys. Rev. E {\bf 89}, 022123 (2014).

\bibitem{Olmi:2014}S. Olmi, A. Navas, S. Boccaletti, and A. Torcini, Phys. Rev. E {\bf 90}, 042905 (2014).

\bibitem{Olmi:2016}S. Olmi and A. Torcini, in {\it Control of Self-Organizing Nonlinear Systems, Springer Series in Energetics},
edited by E. Schoell, S.H.L. Klapp, and P. Hoevel (Springer-Verlag, Berlin, 2016) 

\bibitem{Campa:2015}A. Campa, S. Gupta, and S. Ruffo, J. Stat. Mech.: Theory Exp. P05011 (2015).

\bibitem{Gupta:2012}S. Gupta, M. Potters, and S. Ruffo, Phys. Rev. E
{\bf 85}, 066201 (2012); S. Gupta, A. Campa, and S. Ruffo, Phys. Rev. E {\bf 86}, 061130
(2012). 

\end{thebibliography}
\end{document}